\RequirePackage{fix-cm}
\documentclass[smallextended]{svjour3}       
\smartqed  
\usepackage{amsfonts}
\usepackage[colorlinks,citecolor=blue,linkcolor=blue,urlcolor=blue]{hyperref}
\usepackage{amsmath}
\usepackage{amssymb}
\usepackage{graphicx}
\usepackage{comment}
\usepackage{bm}
\usepackage{mathrsfs}
\usepackage{subfigure}
\usepackage{color}
\usepackage{amsmath,epsfig}
\usepackage{epstopdf}

\renewcommand{\phi}{\varphi}
\renewcommand{\>}{\rangle}

\newcommand{\ket}[1]{|#1\>}

\newcommand{\nn}{\nonumber}
\newcommand{\be}{\begin{equation}}
\newcommand{\ee}{\end{equation}}
\newcommand{\bea}{\begin{eqnarray}}
\newcommand{\eea}{\end{eqnarray}}

\newcommand{\aver}[1]{\left \langle #1\right \rangle}

\newcommand{\ha}{\hat{a}}

\begin{document}

\title{Evaluation of the performance of two state-transfer Hamiltonians in the presence of static disorder
}

\titlerunning{Evaluation of the performance of two state-transfer Hamiltonians}        

\author{A. K. Pavlis        \and
        G. M. Nikolopoulos \and 
        P. Lambropoulos
}


\institute{A. K. Pavlis \and P. Lambropoulos  
\at
Institute of Electronic Structure \& Laser, FORTH, P.O. Box 1385, GR-70013 Heraklion, Greece 
\and
Department of Physics, University of Crete, P.O. Box 2208, GR-71003 Heraklion, Crete, Greece 
\\
\and 
 G. M. Nikolopoulos  
\at
Institute of Electronic Structure \& Laser, FORTH, P.O. Box 1385, GR-70013 Heraklion, Greece 
\\\email{nikolg@iesl.forth.gr}
}

\date{Received: date / Accepted: date}

\maketitle

\begin{abstract}
We analyse the performance of two quantum-state-transfer Hamiltonians in the presence of diagonal and 
off-diagonal disorder, and in terms of different measures. The first Hamiltonian pertains to a fully-engineered chain and the second to a chain with modified boundary couplings. The task is to find which Hamiltonian is the most robust to given levels of disorder and  irrespective of the input state.  In this respect, it is shown that the  performance of  the two protocols are approximately equivalent. 
\end{abstract}

\newpage 

\section{Introduction}
\label{intro}

Over the last decade, the problem of state-transfer and the engineering of quantum networks have evolved into one of the most active  research areas of  quantum technologies. Various faithful state-transfer Hamiltonians (protocols) have been proposed, and networks of various topologies have been analysed \cite{STNE}. 

Among the  proposed Hamiltonians, those with permanently coupled sites have a prominent position as they require no dynamical manipulations for the transfer \cite{STNE}. Such state-transfer  Hamiltonians have the potential of providing  passive quantum networks that perform certain 
communication tasks, without the need of external control (besides the preparation and the read-out of the input state). In that context, the simplest problem one may consider pertains to the faithful (ideally perfect)  transfer of a single-qubit state between the two ends of a spin chain with permanently coupled sites. In analogy to conventional wires used in (opto)electronics,  for the chain to  operate as a passive quantum wire, it has to transfer faithfully {\em any} input qubit state from one end to another at a prescribed time. The transfer is expected to be disturbed by 
 static (or slowly changing) noise, which will be inevitably present in any  physical realization of the spin chain, resulting in diagonal and off-diagonal disorder, pertaining to energies and couplings in the Hamiltonian,  respectively. It is natural to consider that for a given experimental setting the accuracy in the implementation of any given Hamiltonian is known (i.e., the lowest achievable  levels of diagonal and off-diagonal disorder are known), and one would like to know before the implementation, how the various state-transfer  Hamiltonians are expected to perform for the particular levels of disorder, aiming at Hamiltonians which perform reliably for {\em any} input state.  

In this work we  compare the performance of two of the most promising state-transfer Hamiltonians \cite{STNE,spincoupling,PSTQSN,Apollaro,Banchi}, 
in the presence of static diagonal and off-diagonal disorder. Similar studies for the particular protocols in the past have 
been restricted to off-diagonal disorder only, while the performance of the protocols was quantified in terms of  the average-state  fidelity \cite{QSTXXCI}. As mentioned above, for practical reasons we are interested in state-transfer protocols which perform reliably for {\em any} input state. With this objective in mind,  the task here is to reach, if possible, a definitive conclusion about which one of the two Hamiltonians under consideration is the most robust  for given levels of diagonal and off-diagonal disorder. Unfortunately,  for most state-transfer Hamiltonians, including the two discussed here, the quality of the transfer does  depend on the input state; a dependence that is not included in the  average-state fidelity. As has been discussed in the context of a well-studied state-transfer Hamiltonian \cite{Nik}, the predictions of the average-state fidelity have to be taken with  some caution, when the role of the input state is under consideration. The results of Ref. \cite{Nik} clearly suggest that  for the unambiguous quantification of the state-transfer, measures in addition to  
the average-state fidelity  are necessary. 

The two protocols under consideration are outlined in Sec. \ref{sec2}, together with the measures we use throughout this work for the quantification of their performance. Our simulations and the main results are discussed in Sec. \ref{sec3}, followed by conclusions in Sec. \ref{sec4}.

\section{Models and Measures}
\label{sec2}
The two state transfer Hamiltonians under consideration are of the so-called XX spin-chain form  
\be
\mathcal{H}={1\over 2}\sum_{i=1}^{N-1} J_i\left[\sigma^x_i\sigma^x_{i+1}+\sigma^y_i\sigma^y_{i+1}\right]-{1\over 2}\sum_{i=1}^N \varepsilon_i \sigma^z_i,
\ee
where $N$ denotes the number of spin sites, $\sigma_i^{x,y,z}$ are the three Pauli spin operators for the $i$th spin site, $\varepsilon_i$ is the energy of the $i$th spin site and $J_i$ is the exchange interaction between the $i$th and the $(i+1)$th spin site. The XX spin-chain Hamiltonian is isomorphic to a Hubbard  Hamiltonian of the form\cite{WJ,Sachdev}
\be
\mathcal{H}  = -\sum_{i=1}^{N} \varepsilon_i \ha^{\dag}_i \ha_i 
+ \sum_{i=1}^{N-1} J_i ( \ha^{\dag}_i \ha_{i+1} + \ha^{\dag}_{i+1} \ha_i), 
\label{HamHub} 
\ee
The operators $a^{\dag}_i$ $(a_i)$ create (annihilate) a particle at the $i$th site 
with energy $\varepsilon_i$. In this model we can interpret $J_i$ 
as the tunneling coupling between adjacent sites $i$ and  $i+1$. The qubit basis states 
for a single spin (particle) will be denoted by $\lbrace \vert 0\rangle,\vert 1\rangle\rbrace$. 

The first state-transfer Hamiltonian we consider is mirror symmetric and  pertains to the following parameters \cite{STNE,spincoupling,PSTQSN}
\bea
\varepsilon_i=\varepsilon,\quad J_i=J_0^\prime\sqrt{i(N-i)},
\label{protocol1}
\eea
where $J_0^\prime := J_0[N_c(N_c-1)]^{-1/2}\simeq 2J_0/N$ for $N_c = \lceil N/2\rceil\gg 1$.
It allows for ideally perfect transfer of states between the two ends of the chain, and it is isomorphic to 
the rotation of a spin system around the $x$ axis. Hence, we will refer to it as the {\em spin-analogue protocol}. 

The second protocol is also mirror symmetric, but in contrast to the first model it does not require full engineering of the couplings but only of the two outermost couplings i.e., we have  \cite{Apollaro,Banchi}
\bea
\varepsilon_i=\varepsilon,\quad  J_1=J_N=\alpha J_0, 
\quad
\textrm{ and }\quad
 J_i=J_0\quad \textrm{ for }\quad i\neq 1,N, 
\label{protocol2}
\eea
where $\alpha\sim N^{-1/6}$ and $J_0$ is a constant. The two outermost couplings are optimized so that 
the transfer of the state between the two ends of the chain is maximized.  We will refer to this second protocol  as {\em optimal-coupling protocol}. 

For an ideal chain, a Hamiltonian with parameters given by Eq. (\ref{protocol1}) or 
Eq. (\ref{protocol2}), ensure  transfer of an input qubit state 
between the two ends of the chain (the 1st and the $N$th site) at a well-defined time. 
The operation of the Hamiltonians relies on the same principle i.e., the commensurate spectrum \cite{STNE,spincoupling,QSTXXCI}. 
The key difference between the two Hamiltonians is that 
the full spectrum of Hamiltonian (\ref{protocol1}) is commensurate, 
whereas for Hamiltonian (\ref{protocol2}), the spectrum is commensurate 
only in the middle of the energy band, where the overlap of the input state with the corresponding eigenstates is  largest. As a result, irrespective of the measure one may use for the quantification of the transfer,   Hamiltonian (\ref{protocol1}) offers ideally perfect (in the mathematical sense) transfer, whereas Hamiltonian  (\ref{protocol2}) offers  faithful state transfer, in the sense that the transfer is never perfect (e.g.,  the fidelity of the transfer is never strictly equal to 1 even in the ideal scenario).

The  times at which the transfer takes place for the first time, are different for the two protocols. In particular, given that for both of them the maximum coupling 
has been chosen equal to $J_0$, the transfer times are   
 \cite{spincoupling,PSTQSN,Apollaro,Banchi}
\bea
\tau_{s} \simeq \frac{\pi N}{4 J_{0}},\quad {\rm and}\quad \tau_o \simeq 
\frac{0.25 N+0.52N^{1/3}}{J_0}, 
\eea 
for Hamiltonians (\ref{protocol1}) and (\ref{protocol2}), respectively. 
In other words, under these Hamiltonians a  spin chain operates as a passive quantum channel (or wire) for transmission of qubit states. The transfer by means of a spin-analogue Hamiltonian is  
slower than the one for the optimal-coupling Hamiltonian. For the sake of 
simplicity, we may note that  $\tau_o\lesssim  \frac{N}{2 J_0}$ and thus as a rule of thumb we may keep in mind $\tau_s\gtrsim \pi\tau_o/2$. It is worth noting that the specific expressions we give for the transfer times refer to the particular normalization we have adopted throughout our work i.e., $\max_i\{J_i\}= 1$. Nevertheless, the relative magnitude of the two transfer times is independent of the normalization. 

In practise, one expects imperfections in the realization of the Hamiltonians  and the focus of the 
present work is on the the performance of the corresponding quantum channels, in the presence of static disorder.   
The performance of the protocols will be analysed in terms of the fidelity \cite{QCQI}
\be \label{eq:fidelity}
F(\rho,\sigma)\equiv \left[{\rm Tr}\sqrt{\sqrt{\rho}\sigma\sqrt{\rho}}\right]^2,
\ee
where $\sigma$ and $\rho$ are the input and output mixed qubit states.
The performance of a disordered quantum channel depends on the input state whereas, in analogy to 
conventional wires, one is typically interested in quantum channels that operate reliably 
irrespective of the input state (signal). In this spirit, the performance of the two Hamiltonians under 
consideration has to be investigated in terms of measures that are independent of the input state.  
One way to obtain such a measure is to take the minimum of the fidelity (\ref{eq:fidelity}) over all  possible input mixed states $\sigma$ i.e., to consider the worst-case scenario. Employing the joint concavity of the fidelity one can show that it is sufficient to take the minimum over all the possible pure qubit states \cite{QCQI} i.e., 
\be 
F_{\min}=\min_\psi\lbrace F_\psi\rbrace.
\label{Fmin:eq}
\ee
where $\ket{\psi} = \alpha\ket{0}+\beta\ket{1}$ with $\alpha,\beta\in{\mathbb C}$ such that 
$|\alpha|^2+|\beta|^2=1$.

In order to proceed further, we need to obtain an expression for the fidelity $F_\psi$, which refers to 
a single realization of the transfer of an input state $\ket{\psi}$ through the disordered channel. Initially the entire spin chain is in the ground state $\vert \textbf{0}\rangle\equiv\prod_{k=1}^N\otimes\vert 0\rangle_k$ 
\footnote{
Strictly speaking, for the spin-analogue Hamiltonian one does not need the 
assumption for the chain to be initially prepared in the ground (vacuum) state. 
State transfer is expected to take place irrespective of the initial state of the 
chain, provided that the input spin (site) is initially decorrelated from the rest of the chain (e.g., see 
chapter 2 in \cite{STNE}).  The main 
reason is that the evolution operator at the transfer time reduces to a permutation (up perhaps to an unimportant global phase).}. 
The input state $\ket{\psi}$ is prepared at the first spin site and the wavefunction of the entire chain is $\ket{\Psi(0)} = \alpha\ket{\bm 0} + \beta\ket{\bm 1}$, where $\ket{\bm j}\equiv\ket{0_1,\ldots,0_{j-1},1_j,0_{j+1},\ldots,0_N}$ denotes the basis state of the chain with one excitation (spin flip) in the $j$th spin site.  At the transfer time $\tau=\tau_{s(o)}$, the wavefunction of the entire chain has evolved to  
\cite{Nik,Fidref3,QCUSC}
\begin{align} 
\label{psi_tau}
\ket{\Psi(\tau)} =\alpha\vert \textbf{0}\rangle+\beta\sum_{j=1}^N f_{1j}(\tau)\vert {\bm j}\rangle,
\end{align}
with the transition amplitudes  $f_{1j}(\tau)\equiv\langle\textbf{j}\vert e^{-i\mathcal{H}\tau}\vert \textbf{1}\rangle $, while for the sake of simplicity we set $f_{1N}(\tau) = \sqrt{p(\tau)}e^{i\phi}$.
Following a straightforward calculation one can show that the fidelity for input state $\ket{\psi}$ entering Eq. (\ref{Fmin:eq}) is of the form \cite{Nik}
\bea\label{Fpsifidelity}
F_{\psi}(\vert \beta\vert^2,p,\Delta \phi)=&1+\vert \beta\vert^2[-1-p+2\sqrt{p}\cos(\Delta \phi)]
 +[2p-2\sqrt{p}\cos(\Delta \phi)]\vert \beta\vert^4,             
 \nonumber\\               
\eea
with  $\Delta \phi\equiv\phi-\phi_{\rm id}$. 

In the absence of disorder, at the end of an ideal single realization of the transfer, the transition amplitude  would be well-defined i.e., $f_{1N}^{({\rm id})} = \sqrt{p_{\rm id}}e^{{\rm i}\phi_{{\rm id}}}$. 
In general, the probability $p_{\rm id}\leq 1$, where strict equality holds only for perfect state-transfer Hamiltonians (such as the spin-analogue protocol), 
whereas for any faithful state-transfer Hamiltonian (including the optimal-coupling protocol),  $p_{\rm id}$ is close to 1. 
The phase $\phi_{\rm id}$ is determined solely by the energies $\{\varepsilon_i\}$ and the couplings $\{J_i\}$ in the Hamiltonian.  
Hence, for an ideal realization, the phase is fixed and known, and thus 
one can compensate for it at the end of the transfer. 
In the presence of static disorder the energies and the couplings become random 
variables (i.e., they change from realization to realization) and thus the transition amplitude $f_{1N}(\tau)$  becomes a complex random variable as well.  Nevertheless,  without loss of generality we can assume that one still  compensates for $\phi_{\rm id}$ at the end of  a nonideal realization of the transfer, and this is reflected in the phase difference $\Delta\phi$ entering $F_\psi$. 
Equation (\ref{Fpsifidelity}) also shows clearly  the dependence of the fidelity on the input state through its amplitude $\beta$. Taking the minimum over all possible input states, is equivalent to taking the minimum 
of this expression with respect to $|\beta|^2$.  Unfortunately there is no analytic compact solution for  
$F_{\rm min}$ and it has to be calculated numerically for each realization.

An alternative, easy-to-handle state-independent measure is obtained by  averaging Eq. (\ref{Fpsifidelity}) over all possible input states (i.e., over $|\beta|^2$) obtaining the average-state fidelity 
 \cite{Nik,Fidref3,QCUSC,Fidref1,Fidref2}
\be \label{eq:BlochFid}
\bar{F}(p,\Delta \phi)={1\over 2}+{p\over 6}+{\sqrt{p}\cos(\Delta \phi)\over 3}.
\ee

Finally, it is well known that the performance of a quantum channel can be analysed in the framework of its ability to distribute entanglement \cite{QCUSC}. 
More precisely, initially the first site of the chain is isolated from the rest of the chain, and is entangled to another external isolated qubit $1^\prime$ e.g., the two qubits are in the state 
$\ket{\Phi_{1^\prime,1}^+} = (\ket{00}_{1^\prime,1}+\ket{1^\prime,1})/\sqrt{2}$.  
At $t=0$, the interaction between the first and the second qubit of the chain is switched on, and the entire chain thus operates according to one  of  the two state-transfer Hamiltonians described above. 
Hence, the qubit state is transferred from the 1st to the $N$th site of the chain at time  $\tau = \tau_{s(o)}$,  and the interaction between the $N$th site 
and the rest of the chain is switched off.  
If perfect state transfer were possible, the output state would be 
a maximally entangled state between  two distant sites i.e., 
$\ket{\Phi_{1^\prime,N}^+} = (\ket{00}_{1^\prime,N}+\ket{11}_{1^\prime,N})/\sqrt{2}$. 
This is the typical procedure for distributing entanglement between 
distant parties (in the particular case between sites $1^\prime$ and $N$),  which can be used e.g., for teleportation 
(see also Ref. \cite{spincoupling} for an alternative entanglement distribution scheme). 
 For a noisy chain, however, 
one can readily show using Eq. (\ref{psi_tau}), that at the end of the transfer the output state is  
\begin{align}
\rho_{1^\prime,N}(\tau)=&{1\over 2}\left\lbrace(1-\vert f_{1^\prime,N}(\tau_c)\vert^2)\vert 10\rangle\langle 10\vert\right.\nn \\
&\left.+(\vert 00\rangle +f_{1^\prime,N}(\tau)\vert 11\rangle)(\langle 00\vert +f^{*}_{1^\prime,N}(\tau)\langle 11\vert)\right\rbrace,
\end{align}
where for the sake of brevity the labels $1^\prime$ and $N$ have been dropped from the bras and kets.
The entanglement in $\rho_{1^\prime,N}(\tau)$, and thus the performance of the noisy channel, 
can be quantified in terms of the  concurrence \cite{WootPRL98}
\bea
{\cal C} = \left| f_{1^\prime,N}(\tau)\right | = \sqrt{p}, 
\label{conc:eq}
\eea
It is worth noting here that entanglement distribution and state transfer are equivalent with respect to the average-state fidelity (\ref{eq:BlochFid}) \cite{BBpra10}. Here, however, we use the concurrence, which relates to the transfer of probability rather than the average-state fidelity (compare equation \ref{conc:eq} to equation \ref{eq:BlochFid}).

In the following section we analyse the performance of the spin-analogue and the optimal-coupling Hamiltonians  in the presence of  static diagonal and off-diagonal disorder. The performance is 
quantified in terms of the minimum fidelity (\ref{Fmin:eq}) [with $F_\psi$ given by Eq. (\ref{Fpsifidelity})],  the average-state fidelity (\ref{eq:BlochFid}), and the concurrence (\ref{conc:eq}).

\section{Simulations and Results}
\label{sec3}

Throughout this section we assume that the transfer times for the Hamiltonians under investigation 
are much shorter than all the characteristic relaxation times associated with decoherence and dissipation  
mechanisms. We will focus on effects of fabrication  imperfections and slowly varying (with respect to  
$\tau_s$) time-dependent fluctuations in the energies and the couplings  of the Hamiltonians. 
Such imperfections and fluctuations can be described in terms of static diagonal and off-diagonal  disorder in the Hamiltonians i.e., randomness in the energies and the couplings, respectively. 

In view of the different ranges of couplings involved in the two Hamiltonians, we set
\begin{subequations} 
\begin{align}
& \tilde{J}_i\rightarrow \tilde{J}_i(1+\delta_i) \\ 
& \tilde{\varepsilon}_i\rightarrow \tilde{\varepsilon}_i(1+\eta_i),
\end{align} 
\end{subequations}
where $\delta_i$ and $\eta_i$ are uncorrelated Gaussian random variables with zero mean and standard deviations $\sigma_J$ and $\sigma_\varepsilon$, respectively. Note that to facilitate our simulations 
we worked in  dimensionless quantities $\tilde{J}_i=J_i/J_0$ and $\tilde{\varepsilon}_i=\varepsilon/J_0$. For mathematical reasons, we have to consider $\varepsilon \neq 0$. The actual value of $\varepsilon$ does not play any significant role in our work since it contributes a constant term to $\varphi_{\rm id}$ for which, as mentioned above, we compensate at the end of the transfer.

The effects of  static disorder on the performance of a given state-transfer Hamiltonian were analysed in terms 
of independent realizations. More precisely, for a particular state-transfer Hamiltonian and a given  input qubit state  $\ket{\psi}$ we performed a number of realizations, each one simulating the transfer of the particular state through the corresponding  noisy quantum channel. For each realization,  the channel pertained to  a randomly chosen set of couplings and  energies for the Hamiltonian, that were generated as mentioned above, and  remained constant throughout the transfer. At the end of the transfer we kept track of $p$, $\Delta\phi$, $F_\psi$ and $\bar{F}$. Both of $F_\psi$ and $\bar{F}$ 
varied from realization to realization, and by taking an ensemble average over many realizations, the performance of the Hamiltonian was quantified by the ensemble-averaged  quantities $\aver{F_\psi}$ and $\aver{\bar{F}}$, as well as their variances. 
For the estimation of $\aver{F_{{\rm min}}}$ we followed the same procedure but at the end of each realization we calculated $F_{\min}$ as $F_\psi(B,p,\Delta\phi)$, with \cite{Nik}
\be\label{B:eq}
B =
\left\{
  \begin{array}{lr}
   {{1+p-\sqrt{p}\cos(\phi)}\over{4(p-2\sqrt{p}\cos(\phi))}}, &  {\rm if}\, B\in[0,1]\\
    1, & \text{otherwise}.
  \end{array}
\right. 
\ee
This is because, as has been shown in \cite{Nik}, $F_\psi(|\beta|^2,p,\Delta\phi)$ is minimized for 
$|\beta|^2 = B$ i.e,  the state $\ket{1}$ does not always correspond to the worst-case scenario. Hence, for fixed levels of disorder, one has to find the worst-case input-state numerically During the same procedure one can also obtain the ensemble-averaged $\aver{B}$ 
which refers to the average input state that minimizes the fidelity. The case of a perfect classical
channel corresponds to $\aver{p}=1$ and $\aver{\cos(\Delta\phi)}=0$, obtaining from Eq. (\ref{eq:BlochFid}) $\bar{F}_{\rm cl} = 2/3$. 
The classical limit of 2/3 is the best that in principle can be achieved  by a direct projective measurement on the qubit state on one end of the chain, and transfer of the outcome via classical communication to the other end, where the qubit can be prepared according to the outcome.  The limit of $1/2$  corresponds to a random guess of the qubit state $\ket{0}$ or $\ket{1}$, with no measurements \cite{MP95}.

Before presenting our main results, it is worth noting here that the performances of the  spin-analogue and the optimal-coupling protocols have been compared by other authors in the framework of off-diagonal disorder and the ensemble-averaged average-state fidelity $\aver{\bar{F}}$ \cite{QSTXXCI}. 
However, as has been shown in \cite{Nik} for one particular protocol, $\aver{\bar{F}}$ tends to overestimate the performance of protocols, and in some cases it may lead to erroneous conclusions. Hence, in order to reach definitive and reliable conclusions about the performance of the protocols, we consider both diagonal and off-diagonal  disorder, while we compare their performance in terms of different measures.  

 \begin{figure}
   \centering{
   \includegraphics[scale=0.33]{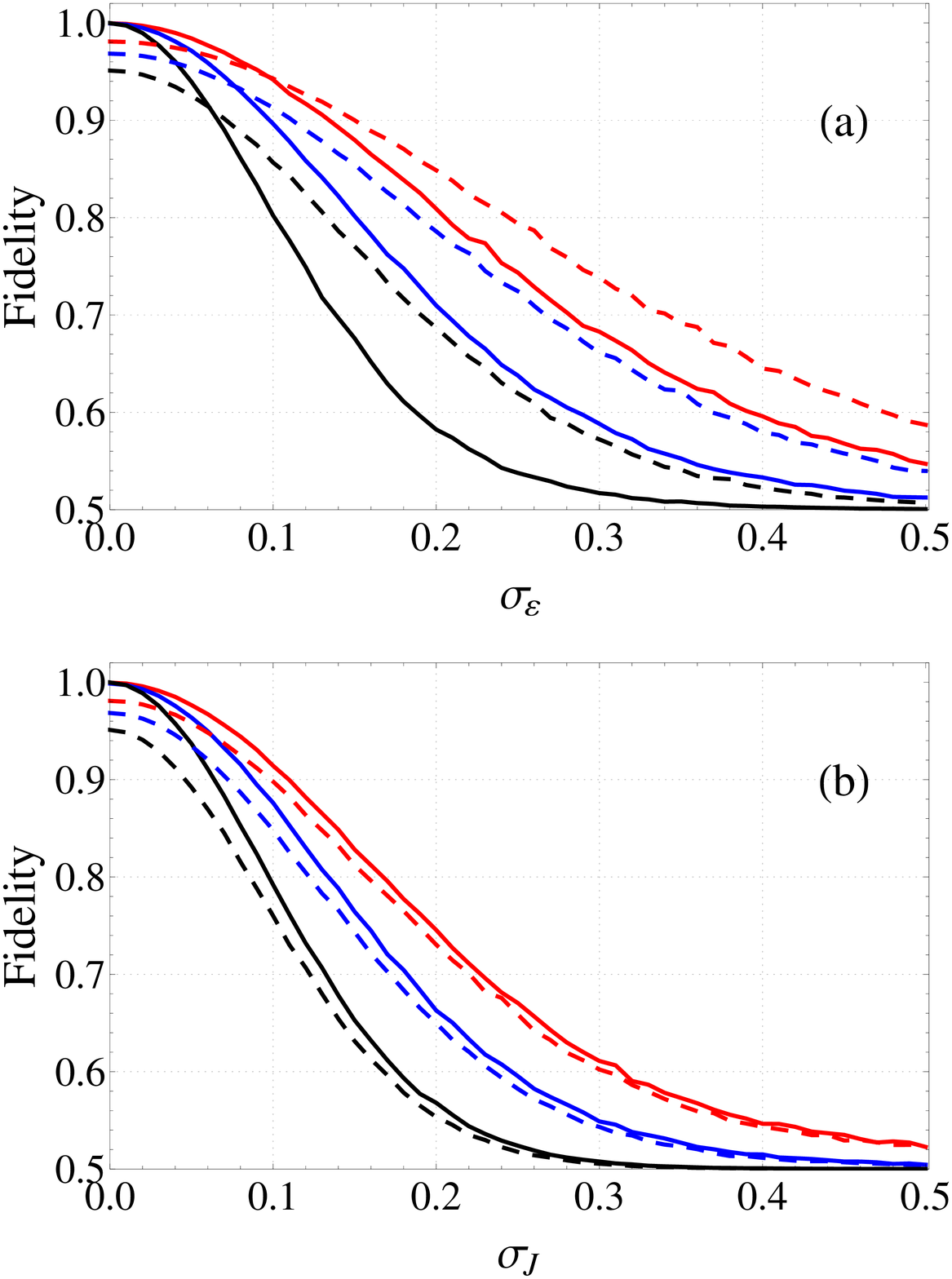}}
  \caption{The ensemble-averaged average-state fidelity $\aver{\bar{F}}$ for the optimal-coupling  protocol (dashed lines) and the spin-analogue protocol (solid lines) as a function of the diagonal (a) and off-diagonal (b) disorder, and for various numbers of (spin) sites. Parameters:  (a) $\sigma_J=0$,  (b) $\sigma_\varepsilon=0$; $N=15$ (red), $N=25$ (blue) and $N=50$ (black). The ensemble average has been obtained on $1000$ independent realizations.}
  \label{fig1}
\end{figure}

\begin{figure}
   \centering
   \includegraphics[scale=0.33]{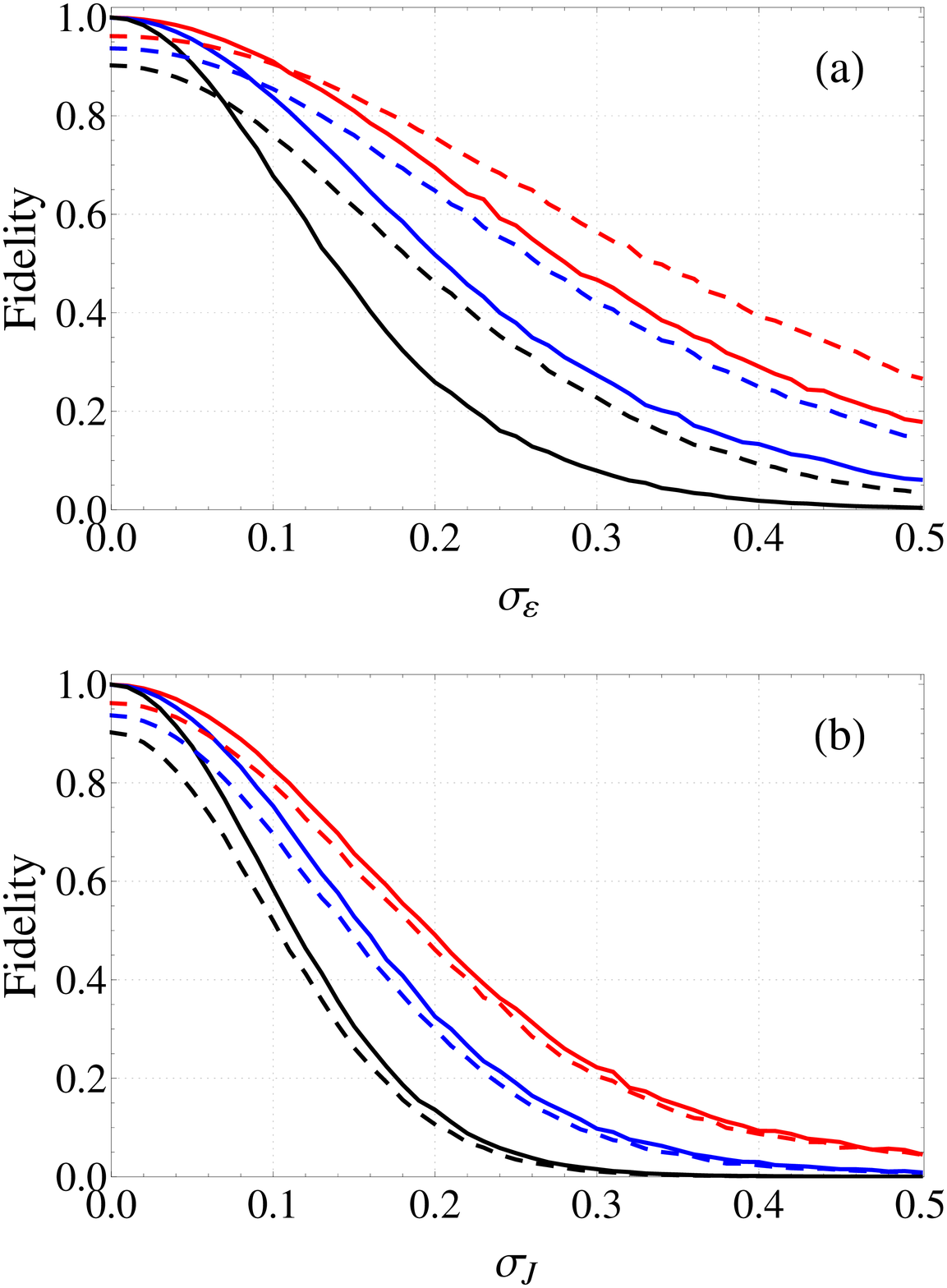}
  \caption{The ensemble-averaged  minimum fidelity $\aver{F_{\min}}$ for the optimal-coupling  protocol (dashed lines) and the spin-analogue protocol (solid lines) as a function of the diagonal (a) and off-diagonal (b) disorder, and for various numbers of (spin) sites. Parameters:  (a) $\sigma_J=0$,  (b) $\sigma_\varepsilon=0$; $N=15$ (red), $N=25$ (blue) and $N=50$ (black). The ensemble average has been obtained on $1000$ independent realizations.}
  \label{fig2}
\end{figure}

\subsection{Diagonal or Off-diagonal disorder}
\label{sec3a}

Let us begin by considering first the performance of the two protocols in the presence of  
one type of disorder only. 
In Fig. \ref{fig1}, $\aver{\bar F}$  is plotted as a function of diagonal and off-diagonal  disorder, for the spin-analogue (solid lines) and the optimal-coupling (dashed lines)  Hamiltonians, and for 
chains with various numbers of spin sites. As  expected,  in all of the cases $\aver{\bar{F}}$ drops with increasing disorder. Our studies show that for either of the two protocols and for any $N$, this behaviour is well approximated by a Gaussian of the form 
\bea
G(\sigma_\varepsilon,\sigma_J) =
A\exp\left ( -cN \sigma_J^2- dN \sigma_\varepsilon^2\right )+C,
\label{fit:eq}
\eea 
to first order in $N$,  
where $A,C,c$ and $d$ are fitting parameters that depend on the protocol under consideration. More precisely, for the spin-analogue Hamiltonian  $A=C=1/2$, $c\simeq 1.07$ and $d\simeq 0.7$, while for the optimal-coupling protocol $A=p^2/3+ p/6$, $C=1/2$,  $c\simeq 1.2$ and $d\simeq 0.46 $. The same scaling law for the spin-analogue Hamiltonian has been  obtained   in Ref. \cite{De Chiara}, 
and here we see that the  law applies to the optimal-coupling Hamiltonian as well, albeit with different parameters. Moreover, 
we see that the two protocols are practically equivalent with respect to their robustness against 
off-diagonal disorder, but  their performance is quite different with respect to diagonal disorder. 
Indeed we find that the optimal-coupling protocol is more robust than the spin-analogue Hamiltonian against 
diagonal disorder with $\sigma_{\varepsilon}<0.5$, and this difference becomes more pronounced for  increasing number of (spin) sites in the chain.  Finally, for a broad range of standard deviations, 
off-diagonal disorder seems to be more catastrophic than diagonal disorder, for both protocols.

\begin{figure}[t!]
   \centering
   \includegraphics[scale=0.33]{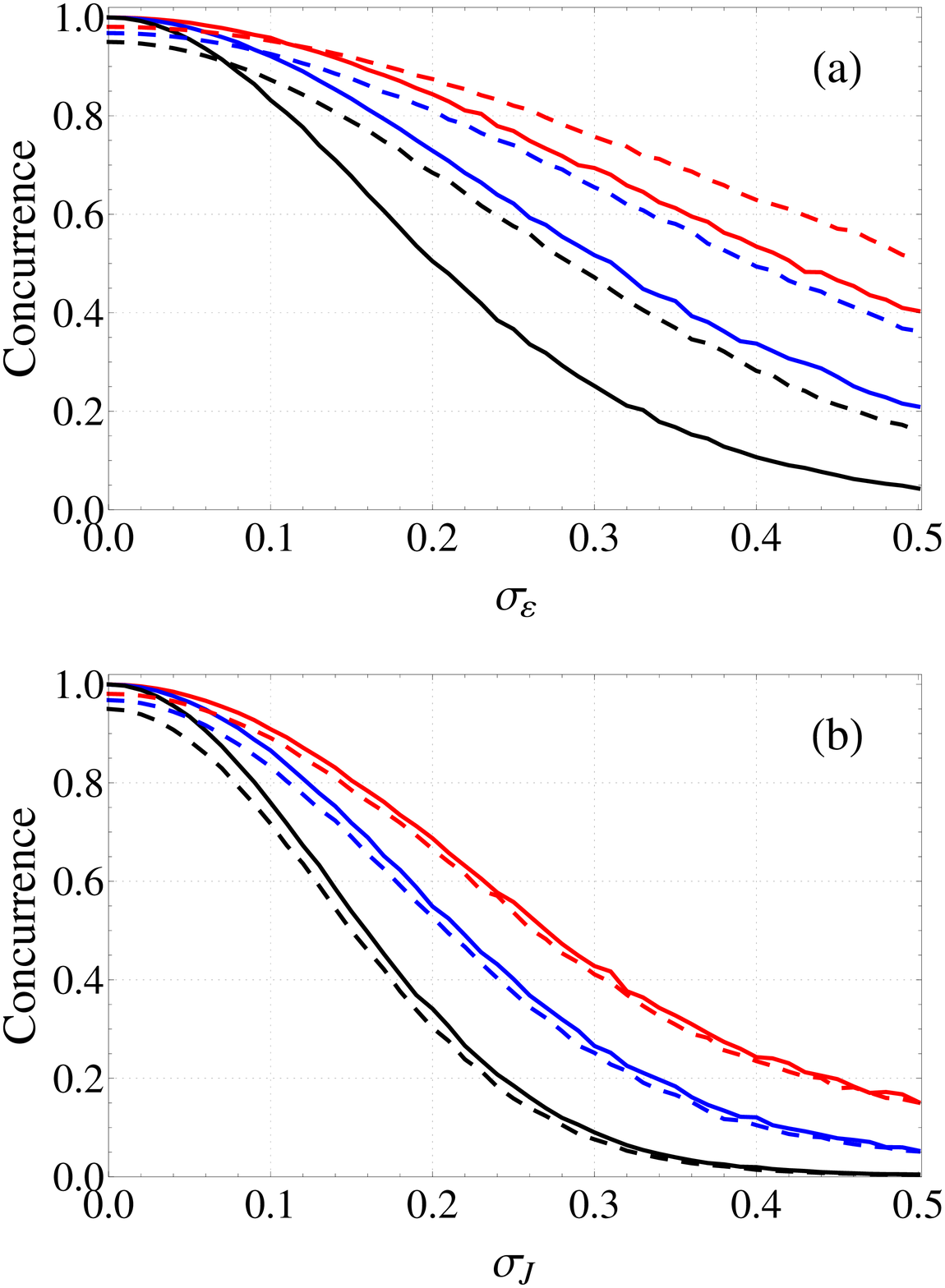}
  \caption{
  The ensemble-averaged concurrence $\aver{\cal C}$ for the optimal-coupling  protocol (dashed lines) and the spin-analogue protocol (solid lines) as a function of the diagonal (a) and off-diagonal (b) disorder, and for various numbers of (spin) sites. Parameters:  (a) $\sigma_J=0$,  (b) $\sigma_\varepsilon=0$; $N=15$ (red), $N=25$ (blue) and $N=50$ (black). The ensemble average has been obtained on $1000$ independent realizations.}
  \label{fig3}
\end{figure}

In Fig. \ref{fig2}, the performance of the two protocols in the presence of either of the two 
types of disorder is analysed in terms of the ensemble-averaged minimum fidelity $\aver{F_{\min}}$. 
Note the different scales when comparing to 
Fig. \ref{fig1}, since by contrast to $\aver{\bar{F}}$ which is averaged over all possible input states, $\aver{F_{\min}}$ refers to the worst-case scenario and thus approaches zero for large disorder. 
As in the case of $\aver{\bar{F}}$, $\aver{F_{\min}}$ also follows a Gaussian law with respect to disorder. 
In particular we have found that a rather good fit to our numerical results is provided by (\ref{fit:eq}) with the following parameters.  For the spin-analogue protocol $A=\aver{B}$, $C=1-\aver{B}$, $c\simeq 1.07$ and $d\simeq 0.8$ while for the optimal-coupling protocol $A=\aver{B} \left( 2 p^2/ 3+p/3\right )$, $c\simeq 1.2$,  
$d\simeq 0.40 $ and $C$ as before. Comparing to the fit parameters we obtained for the data on $\aver{\bar{F}}$, we see that for both protocols the decay parameters $c$ and $d$ are quite close to those 
for the fit to $\aver{\bar{F}}$.  So, as long as one focuses on very weak disorder $\sigma_{\varepsilon},\sigma_J\lesssim 0.1$, the two measures predict similar performance for the protocols. For stronger disorders, however, the predictions of the 
two measures begin deviating considerably, with the average-state fidelity $\aver{\bar{F}}$  approaching $1/2$ smoothly, while $\aver{F_{\min}}$  keeps dropping below $1/2$, according to the Gaussian law mentioned above.

When the performance of the protocols is analysed in terms of the the average concurrence, 
the main observations are the same (see Fig. \ref{fig3}). 
In fact the concurrence seems to  drop slower than both 
$\aver{\bar{F}}$  and $\aver{F_{\min}}$, which suggests that the fidelities are much stricter 
measures. This is because as seen by Eq. (\ref{conc:eq}), the concurrence does not depend on the 
phase, but only on the transition amplitude (see also related discussion in \cite{QCUSC}).

Having gained some insight on the performance of the the protocols when only one of the disorders 
is present, in the following section we discuss the performance of the protocols in the presence of both disorders


\begin{figure*}[t]
   \centering
   \includegraphics[width=0.9\textwidth]{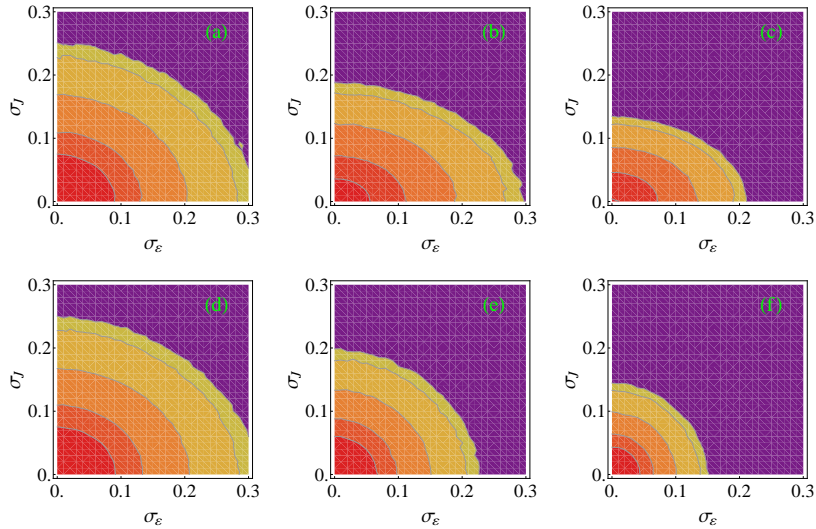}
  \caption{The ensemble-averaged average-state fidelity $\aver{\bar{F}}$ for the optimal-coupling (a,b,c) and the spin-analogue protocol  
  (d,e,f), in the presence of both diagonal and off-diagonal disorder, and for $N=15$ (a,d),
   $N=25$ (b,e) and for $N=50$ (c,f) sites.  The contours are for $\aver{\bar{F}} = \{0.95,0.9,0.8,0.7,0.67\}$ and the ensemble averages have been obtained over 1000 independent realizations.}
  \label{fig4}
\end{figure*}

\subsection{Diagonal and Off-diagonal disorder}
\label{sec4}

The performance of the two protocols with respect to the average-state fidelity $\aver{\bar{F}}$ 
is shown in Fig. \ref{fig4}, for various numbers of sites.  The two models are practically 
equivalent for moderate number of sites ($N\lesssim 20$), while the optimal-coupling scheme 
seems to gain ground for larger number of sites. The optimal-coupling scheme is more sensitive to 
off-diagonal disorder than to diagonal, whereas the performance of the spin-analogue scheme does not exhibit so pronounced asymmetry. 

\begin{figure}[t!]
   \centering
   \includegraphics[width=0.9\textwidth]{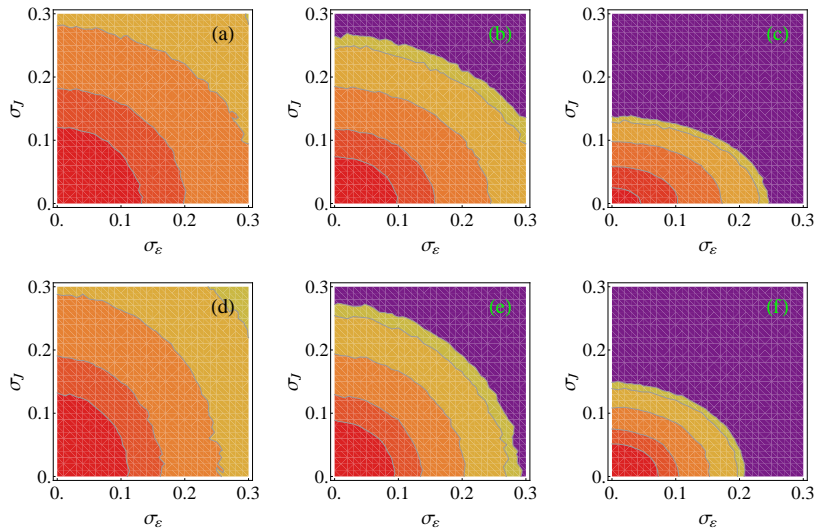}
  \caption{The ensemble-averaged fidelity $\aver{F_{\psi}}$ for the optimal-coupling (a,b,c) and the spin-analogue protocol  
  (d,e,f), in the presence of both diagonal and off-diagonal disorder, and for $N=15$.  
The density plots on the top refer to the optimal-coupling protocol, and the ones at the bottom to the spin-analogue protocol.  The input state is   (a,d) $\vert \beta\vert^2=0.4$; (b,e) $\vert \beta\vert^2=0.6$; (c,f) $\vert \beta\vert^2 = B$.
      The contours are for $\aver{F_{\psi}}= \{0.95,0.9,0.8,0.7,0.67\}$ and the ensemble averages  have been obtained over 1000 independent realizations. 
}
  \label{fig5}
\end{figure}

\begin{figure}[t!]
   \centering
   \includegraphics[width=0.9\textwidth]{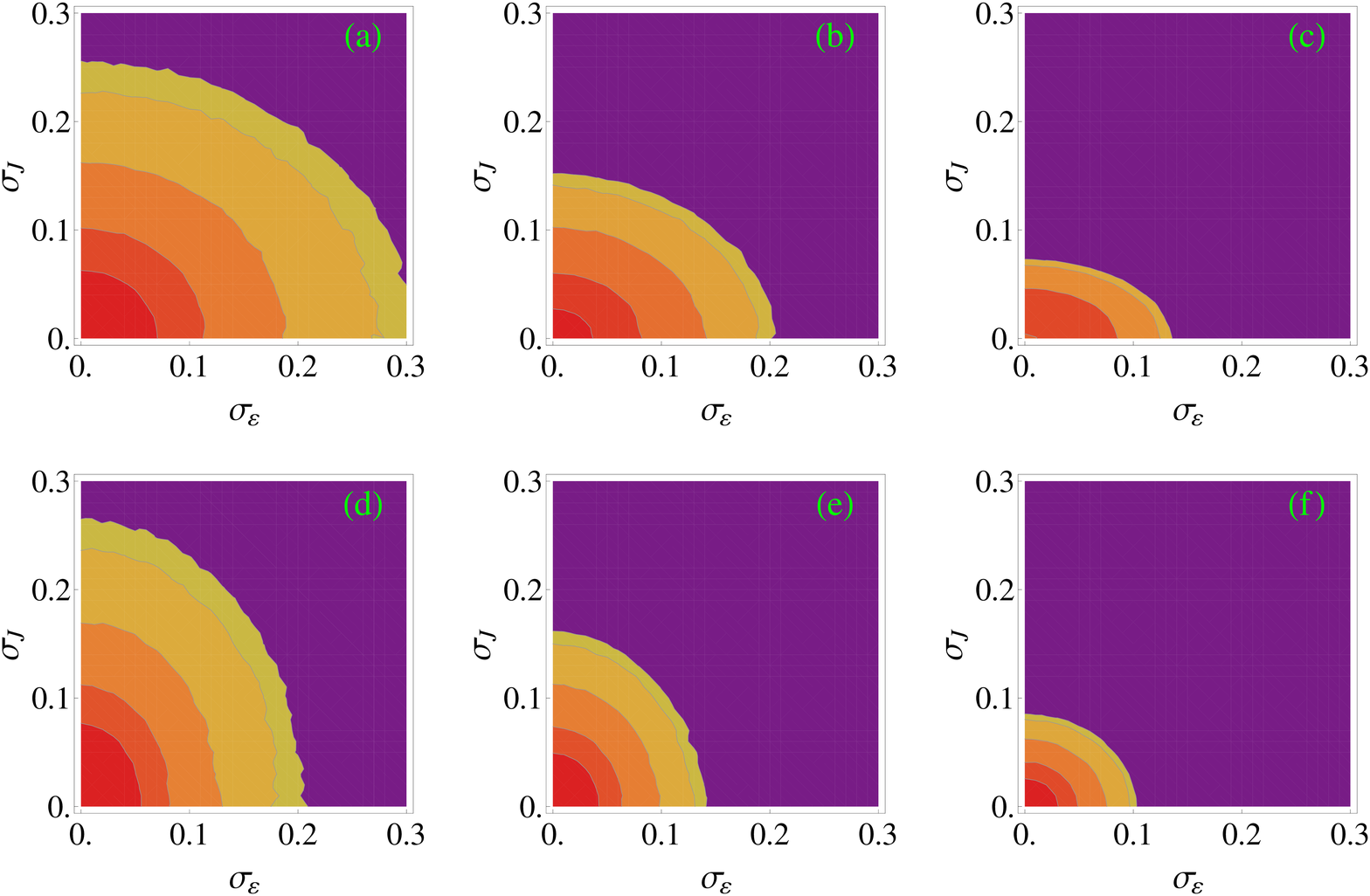}
  \caption{The ensemble-averaged fidelity $\aver{F_{\psi}}$ for the optimal-coupling (a,b,c) and the spin-analogue protocol  
  (d,e,f), in the presence of both diagonal and off-diagonal disorder, and for $N=50$.  
The density plots on the top refer to the optimal-coupling protocol, and the ones at the bottom to the spin-analogue protocol.  The input state is   (a,d) $\vert \beta\vert^2=0.4$; (b,e) $\vert \beta\vert^2=0.6$; (c,f) $\vert \beta\vert^2 = B$.
      The contours are for $\aver{F_{\psi}} = \{0.95,0.9,0.8,0.7,0.67\}$ and the ensemble averages have been obtained over 1000 independent realizations.}
  \label{fig6}
\end{figure}

The performance of the two schemes for various input states is plotted in 
Figs. \ref{fig5},  and \ref{fig6} for two different numbers of sites.  
Recall here that for each realization, the minimum fidelity $F_{\min}$ corresponds to $F_\psi$ estimated for 
$|\beta|^2=B$, with $B$ given by Eq. (\ref{B:eq}).
We  see  clearly that the quality of the transfer is totally dependent on the input state. For states with  $\vert\beta\vert^2\in [0,0.4]$ the ensemble-averaged fidelity $\aver{F_\psi}$  does not drop below $2/3$ even for disorders  $\sigma_\varepsilon,\sigma_J \simeq 0.3$. On the contrary for $\vert\beta\vert^2\in [0.6,B]$ the transfer starts to fail for both protocols. For $\vert\beta\vert^2 = 0.6$ the protocols fail for large disorder 
i.e., for $\sigma_\varepsilon,\sigma_J \gtrsim 0.25$,  and the region of failure becomes larger 
as we increase  $\vert\beta\vert^2$ approaching the worst-case scenario corresponding to 
$\vert\beta\vert^2=B$. Once more the optimal-coupling scheme exhibits the same asymmetry with respect to the two types of disorder,  while for the spin-analogue coupling the asymmetry is not so pronounced.  

An obvious conclusion from Figs. \ref{fig4}-\ref{fig6} is that $\aver{\bar{F}}$ tends to  underestimate the state transfer for input states with $\vert\beta\vert^2\in [0,0.4]$ and to overestimate it for $\vert\beta\vert^2\in[0.6,1.0]$. Hence, in general, particular caution is necessary when $\aver{\bar{F}}$ is employed as a measure for the  quality of the transfer. In Ref. \cite{Nik} the ensemble-averaged minimum fidelity  $\aver{F_{\min}}$ has been proposed as a more reliable measure since it is also state independent and  provides a lower bound for the fidelity of the transfer of any input qubit state. Although the lower bound can be rather loose, as we see from Figs. \ref{fig5} and \ref{fig6}, it never overestimates the performance of the protocol under consideration. 

\begin{figure}
   \centering
   \includegraphics[scale=0.22]{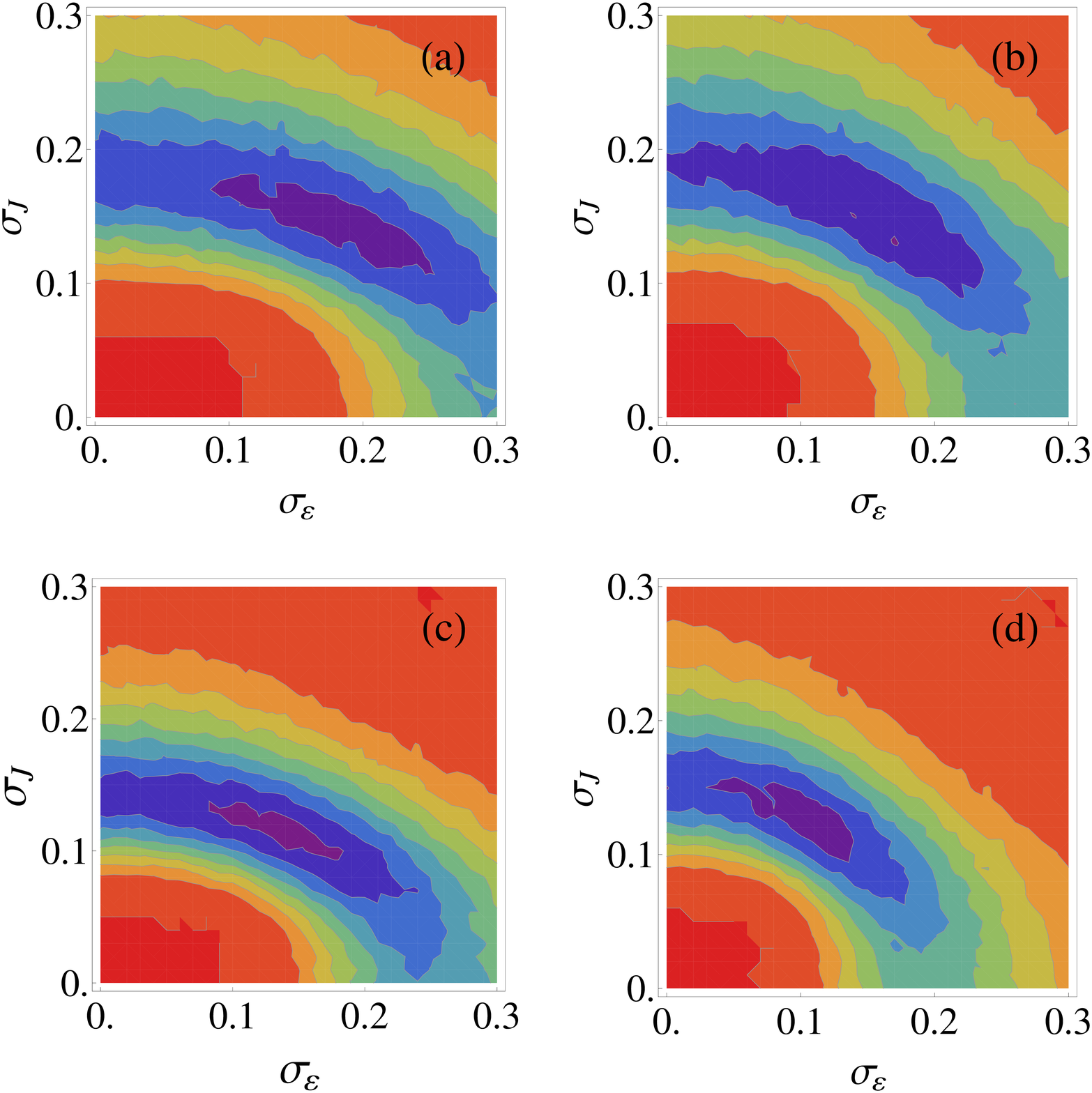}
   \includegraphics[scale=0.38]{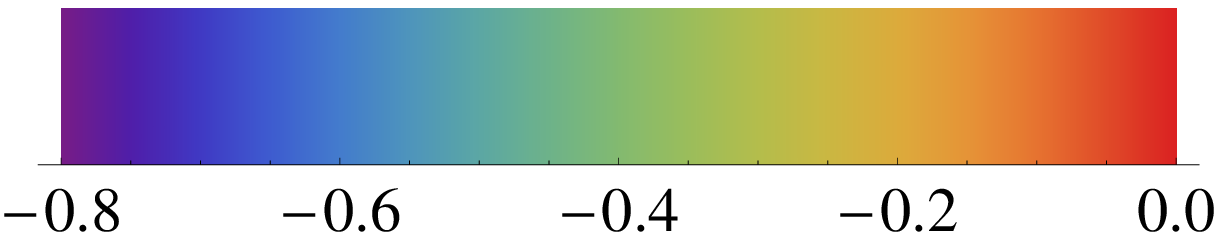}
  \caption{The difference ${\rm Pr}(\aver{F_{\min}}>\bar{F}_{\rm cl} )-{\rm Pr}(\aver{\bar{F}}>\bar{F}_{\rm cl} )$ for the  optimal-coupling protocol (a,c) and the spin-analogue coupling (b,d) for number of two different numbers of sites:  $N=15$ (a,b) and $N=25$ (c,d).}
  \label{fig7}
\end{figure}

Finally,  we  have looked at another  interesting quantity i.e.,  the probability of successful transfer.  
As before, a number of independent realizations were performed, and for each one of them, the disorder was chosen at random and was fixed during the transfer.  We performed such simulations for both protocols, keeping track of the cases where either $\aver{F_{\min}}$ or $\aver{\bar{F}}$ dropped below the classical threshold $\bar{F}_{\rm cl}$. In Fig. \ref{fig7} we plot  the difference of the estimated  probabilities ${\rm Pr}(\aver{F_{\min}}>\bar{F}_{\rm cl} )-{\rm Pr}(\aver{\bar{F}}>\bar{F}_{\rm cl} )$. One sees immediately that  $\aver{\bar{F}}$ always overestimates the performance of both protocols.  Even for moderate disorder ($\sigma_{J},\sigma_{\varepsilon} \sim 0.2 $), this overestimation is of the order of 0.5, whereas it drops to about 0.1, only for  ($\sigma_{J},\sigma_{\varepsilon} \lesssim 0.1 $).  Moreover, we see once more that the two protocols are practically equivalent, as differences are not either extensive or pronounced.


\section{Discussion and Concluding Remarks}
\label{sec4}

We have analysed the performance of two state-transfer protocols in the presence of static diagonal and off-diagonal disorder. The first protocol pertains to fully engineered chains (spin-analogue protocol) whereas the second one to modified boundary couplings (optimal-coupling protocol). For given levels of disorder, we  investigated which of the two protocols is the most robust,  irrespective of the input state. To this end, we employed different measures for the quantification of the transfer, including the 
{ensemble-averaged minimum fidelity $\aver{F_{\min}}$, and   average-state fidelity $\aver{\bar{F}}$.  
The latter has not been found very useful for our purposes, since it fails to  describe accurately the performance of the protocols for a large class of input states, and it may lead to faulty conclusions about the success of the transfer. On the other hand,  the minimum fidelity provides  a 
lower bound for the quality of transfer. Although in general this bound is not tight and the quality of transfer can be considerably higher for some input states, it never overestimates the performance of the protocol under consideration as it refers to the ``worst-case scenario".

Our results suggest that the two protocols are practically equivalent, albeit with some minor differences in their performance mainly with respect to 
diagonal disorder. That is, the optimal-coupling protocol appears to be less susceptible to diagonal disorder than the spin-analogue coupling.  
Moreover, both protocols exhibit an asymmetry with respect to their sensitivity to diagonal and off-diagonal disorder, but for the optimal-coupling protocol this asymmetry is more pronounced. The strong similarities of the two protocols can be attributed to the fact that their operation relies on the same principle i.e., the commensurate spectrum 
for the eigenstates involved in the evolution, whereas their minor differences to the fact that, in contrast to the spin-analogue protocol, the optimal-coupling protocol does not pertain to a fully commensurate spectrum. It is  worth pointing out here  that the aforementioned asymmetry is also evident when one looks at the disturbance of the commensurate spectra by the diagonal and the off-diagonal disorder independently. This fact strongly supports the fundamental role of the commensurate spectra in the operation of the two protocols. 

As was expected, for both protocols and for fixed level of disorder, the fidelities drops with increasing number of sites in the chain. 
This is because the transfer time increases (almost linearly) with $N$, and thus the system experiences the effects of disorder for longer periods.  
For  small spin chains ($N\lesssim 15$) the two 
protocols are almost equivalent, while for a moderate number of spin sites ($N\sim 20$) the optimal-coupling protocol appears to be more robust, mainly because of the aforementioned asymmetry with respect to diagonal and off-diagonal disorder. 
For larger spin chains $N>30$, the differences of the two protocols will begin decreasing again until they become negligible due to the rapid decrease of fidelity with $N$. 

The present results have been presented in a generic framework so as to provide a benchmark case and guide for the implementation of quantum-state-transfer schemes in various experimental platforms (see \cite{exp1,exp2} for a comprehensive summary of related experiments). 
Although the present analysis has been in terms of relative errors in the diagonal and off-diagonal terms of the Hamiltonian, if necessary, our performance plots can be converted to plots in terms of absolute errors, 
by rescaling accordingly the depicted standard deviations $\sigma_{\varepsilon}$ and $\sigma_{J}$.

\end{document}